\documentclass[preprint,amsmath,amssymb,aps,prl,longbibliography,lengthcheck]{revtex4-1}

\usepackage{graphicx}
\usepackage{dcolumn}
\usepackage{bm}
\usepackage{hyperref}

\begin{document}

\preprint{APS/123-QED}

\title{Nonlinear dynamics of beta induced Alfv\'en eigenmode driven by energetic particles}

\author{X. Wang$^1$, S. Briguglio$^2$, L. Chen$^{1,3}$, G. Fogaccia$^2$, C. Di Troia$^2$, G. Vlad$^2$, F. Zonca$^{2,1}$}

\affiliation{~$^1$Institute for Fusion Theory and Simulation, Zhejiang University, Hangzhou 310027, P.R.China}
\affiliation{~$^2$Associazione Euratom-ENEA sulla Fusione, C.P. 65 - I-00044 - Frascati, Italy}
\affiliation{~$^3$Department of Physics and Astronomy,  University of California, Irvine CA 92697-4575, U.S.A.}





\begin{abstract}
Nonlinear saturation of beta induced Alfv\'en eigenmode, driven by slowing down energetic particles via transit resonance, is investigated by the nonlinear hybrid magnetohyrodynamic gyro-kinetic code (XHMGC). Saturation is characterized by frequency chirping and symmetry breaking between co- and counter-passing particles, which can be understood as the the evidence of resonance-detuning. The scaling of the saturation amplitude with the growth rate is also demonstrated to be consistent with radial resonance detuning due to the radial non-uniformity and mode structure. 
\end{abstract}

\pacs{Valid PACS appear here}

\maketitle

Apart from Toroidal Alfv\'en Eigenmodes~\cite{cheng85} (TAE), the existence of Beta induced Alfv\'en Eigenmodes~\cite{heidbrink93} (BAE) in the kinetic thermal ion (KTI)~\cite{zonca96a,chen07} gap of the shear Alfv\'en wave (SAW) continuous spectrum is also widely recognized to be a concern for the good confinement of energetic particles (EP) in fusion plasmas~\cite{heidbrink93,heidbrink95}.  BAEs are particularly important in the study of low frequency fluctuations of the shear Alfv\'en wave (SAW) spectrum, since they can be excited by both fast ions (at long wavelengths) as well as by thermal ions (at short wavelengths)~\cite{zonca99,nazikian06}. In addition to Alfv\'en eigenmodes (AE), which are normal modes of the thermal plasma, strongly driven energetic particle continuum modes~\cite{chen94} (EPM) may be also excited at the characteristic frequencies of the EPs, in the presence of a sufficiently intense EP free energy source.

The different nature of AEs and EPMs suggests the existence of two regimes in the nonlinear dynamic evolution of a single toroidal mode number ($n$) coherent SAW driven by EPs. Near marginal stability, when the system is weakly driven and EPMs are not excited, the nonlinear dynamics is the same as in a uniform system and wave saturation can occur when wave-particle trapping flattens the particle distribution in the resonance region~\cite{berk90}. Meanwhile, other physical mechanisms, such as Compton scattering off the thermal ions~\cite{hahm95} and mode-mode couplings, enhancing the interaction with the SAW continuous spectrum by nonlinear frequency shift~\cite{zonca95,chen98}, can also be important, depending on the parameter regimes. All these phenomena are local and independent of the radial mode structure. However, when the system is strongly driven and EPMs~\cite{chen94} are excited at the EP characteristic frequencies, strong EP transport occurs in avalanches~\cite{zonca05}. This phenomenology is strictly related with the resonant character of the modes, which tend to be radially localized where the drive is strongest, and with global readjustments in the EP radial profiles. In the strong nonlinear regime, local and global effects must be treated at the same footing in order to capture the crucial role of radial non-uniformity. 

In this letter, we investigate the transition regime in between the two limiting conditions described above, in order to illustrate the effect of equilibrium non-uniformity as the mode drive is increased above marginal stability. In particular, we analyze AEs resonant excitation below the EPM threshold and choose BAE as specific case because: (1) rich phenomena are observed in experiments~\cite{heidbrink93,heidbrink95}; (2) a single poloidal harmonic is dominant; (3) the mode is very localized, so that finite mode width effects and finite interaction length are easily illustrated. Here, we focus on EP nonlinear dynamic behaviors and their consequences on BAE saturation. Thermal ion nonlinear dynamics may also be important, but it will be analyzed in a separate paper.

In our work, BAE driven by EPs are studied with the extended version of nonlinear magnetohydrodynamic (MHD)-Gyrokinetic code HMGC~\cite{briguglio95} (XHMGC). These extensions include both thermal ion compressibility and diamagnetic effects~\cite{wang11}, in order to account for thermal ion collisionless response to low-frequency Alfv\'enic modes driven by EPs (e.g., kinetic BAE (KBAE)~\cite{wang10}) and finite parallel electric field due to parallel thermal electron pressure gradient. Previous linear simulation studies show that BAE/KBAE can be destabilized by EPs via resonant wave particle interactions~\cite{zhang10,wang10}. Here, we ignore diamagnetic effects of kinetic thermal ions by assuming uniform thermal ion density and temperature profiles, but keep kinetic thermal ion compression effects, in order to correctly describe the KTI gap.  Meanwhile, we consider nonlinear wave-particle interactions with EPs, but neglect both mode-mode coupling effects and nonlinear kinetic thermal ion response. The purpose is to isolate the nonlinear physics due to EPs, which do play dominant roles for sufficiently strong drive. 

\begin{figure}[t]
\includegraphics[width=90mm,height=65mm]{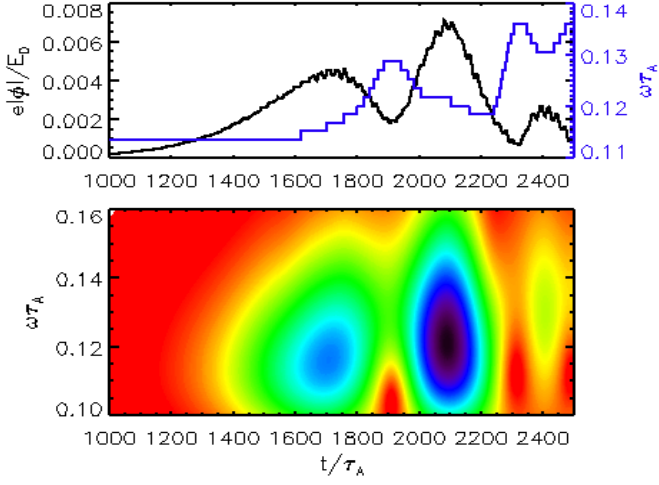}
\caption{\label{fig:omega} Upper panel shows the amplitude evolution (black line) and frequency evolution of the mode (blue line), given by the maximum in the wavelet transform spectrum. Lower panel shows the wavelet transform spectrum for the frequency evolution.}
\end{figure}

We investigate a BAE mode localized around the rational surface at $(r/a)\approx 0.5$ with safety factor $q=2$, dominated by toroidal mode number $n=2$ and poloidal mode number 
$m=4$ harmonic. The BAE-SAW continuum accumulation point frequency can be calculated from kinetic theory~\cite{zonca96a} and is given by $\omega_{cap}=0.127/\tau_A$, where $\tau_A=R_0/V_A$ is the Alfv\'en time, with $V_A$ the Alfv\'en speed on axis and $R_0$ the major radius. The mode is driven by EPs with an isotropic slowing-down distribution and characteristic transit frequency at the birth energy $\omega_{tmax}=(2E_0/m_H)^{1/2}/(qR_0)\simeq0.2/\tau_A$. The mode is excited at $\omega_0\simeq 0.114/\tau_A$, within the KTI gap, and the linear growth rate is $\gamma_L\simeq 0.006/\tau_A$. As the mode grows and saturates, the mode frequency is chirping up quickly, as shown in Fig.~\ref{fig:omega}, because of EP radial redistribution and the corresponding nonlinear change in the BAE dielectric response. To show this, we use the general fishbone like dispersion relation, $i\Lambda(\omega)=\delta W_f+\delta W_k$~\cite{chen07,zonca99}, with the generalized inertia term $i\Lambda\simeq-\sqrt{\omega^2_{cap}-\omega^2}/\omega_A$ for BAE~\cite{zonca96a}, $\omega_A=v_A/(qR_0)$ and $v_A$ being the Alfv\'en frequency and speed, respectively, and $\delta W_f$ and $\delta W_k$ representing fluid and kinetic contributions to the potential energy~\cite{chen07,zonca99}. This allows us to write the nonlinear frequency shift as
\begin{eqnarray}\label{eq:fldr}
\left(\frac{\omega-\omega_0}{\omega_A}\right)\bigg [ \frac{\omega_0}{\sqrt{\omega^2_{cap}-\omega_0^2}}
-\omega_A\frac{\partial}{\partial\omega_0}\mathbb{R}e\delta W_{kL}\nonumber\\
-i\omega_A\frac{\partial}{\partial\omega_0}\mathbb{I}m\delta W_{kL}\bigg ]=\mathbb{R}e\delta W_{kNL}+i\mathbb{I}m\delta W_{kNL} \ \ ,
\end{eqnarray}
having separated linear (L) and nonlinear (NL) contributions to the potential energy. Equation~(\ref{eq:fldr}) shows that upward frequency chirping is expected, due to EP radial redistributions $(\mathbb{R}e\delta W_{kNL}>0, \mathbb{I}m\delta W_{kNL}<0$~\cite{chen07,zonca99}). This result still holds when $|\mathbb{R}e\delta W_{kNL}|\ll|\mathbb{I}m\delta W_{kNL}|$, since -- in the present case of slowing down EPs -- $\partial_{\omega_0}\mathbb{I}m\delta W_{kL}>0$.
\begin{figure}[t]
\includegraphics[width=75mm,height=70mm]{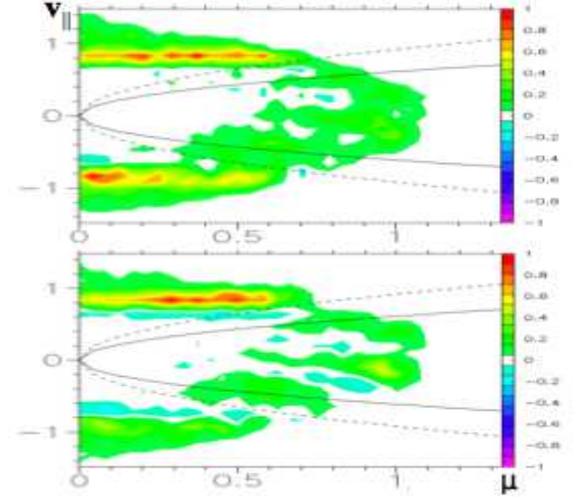}
\caption{\label{fig:power} Wave-particle power exchange in the EP $(\mu, v_\parallel)$ plane during linear phase (upper panel) and saturation phase (lower panel), where $\mu$ is the magnetic moment ($v_\parallel$ and $\mu$ are normalized w.r.t. $\sqrt{E_0/m_H}$ and $E_0/\Omega_0$ with $\Omega_0=eB_0/(m_Hc)$, respectively). The positive sign in the color bar corresponds to ions pumping energy into the wave. Solid and dashed lines are passing-trapped boundaries, calculated for the inner and outer limits of the mode radial width.}
\end{figure}

\begin{figure}[t]
\includegraphics[width=70mm,height=54mm]{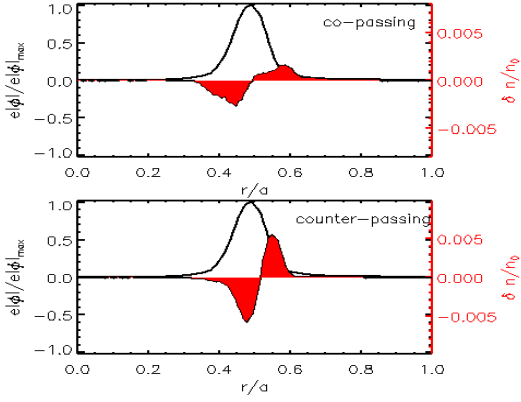}
\caption{\label{fig:density} Radial dependence of the magnetic flux averaged EP fluctuating distribution, integrated over $\mu$ and on a narrow interval $\Delta v_\parallel$ near resonance. Upper panel shows the result for co-passing particles. Lower panel shows the result for counter-passing particles. The radial mode structure with $n=2$ and $m=4$ is shown in the background for reference.}
\end{figure}

Due to the preferred direction in mode frequency chirping, the symmetry between co-passing ($v_\parallel>0$) and counter-passing ($v_\parallel<0$) particles is also broken, as visible in Fig.~\ref{fig:power}, showing the wave-particle power exchange in both linear (upper panel) and saturation phase (lower panel). Results show that the interaction is predominantly driven by transit resonance. Both co-passing and counter-passing  particles are giving energies to the wave in the linear phase, with almost symmetric response in $v_\parallel$ space. 
Different fine structures in the resonant region are due to different orbits of co- and counter-passing particles. However, in the saturation phase, co-passing particles are still driving the mode, while counter-passing particles do not. This is evidence of resonance detuning, explained as follows. The passing particle resonance condition is
\begin{equation}\label{eq:rc}
\omega=\ell\omega_t+(n\bar{q}-m)\sigma\omega_t\ \ ,
\end{equation}
where $\sigma=sgn(v_\parallel)$, $\bar{q}$ is the orbit average of $q$, $\bar{q}=(2\pi)^{-1}\oint qd\theta$ and the transit frequency is defined as $\omega_t=2\pi/\oint d\theta/\dot{\theta}$. Here we have used Clebsch toroidal flux coordinates $(\psi, \theta, \xi)$, with the equilibrium magnetic field ${\bf B}=\nabla\psi\times\nabla\xi$, $\psi$ the magnetic flux and the poloidal angle $\theta$ describing the position along ${\bf B}$. The resonance condition, Eq.~(\ref{eq:rc}), can be expressed as $\dot{\Theta}=0$; i.e., stationarity in the wave-particle phase. For the transit resonance under investigation, we can define the resonance detuning as 
\begin{equation}\label{eq:detuning}
\Delta\dot{\Theta}\simeq(n\bar{q}_r\sigma+\frac{\omega_0}{\omega^2_t}\partial_r\omega_t)\omega_t\Delta r-(\omega-\omega_0)\ \ ,
\end{equation}
where $\bar{q}_r=d\bar{q}/dr$ and $\omega_0=\ell\omega_t+(n\bar{q}-m)\sigma\omega_t$. In the present condition, the resonance-detuning, due to radial particle displacement, is dominated by the first term in parenthesis. Thus, for $\dot{\omega}>0$ (upward frequency chirping),  particles that are transported out, while transferring energy to the wave, more easily maintain the resonance condition for $v_\parallel>0$ (co-passing) than $v_\parallel<0$ (counter-passing). This explains the results of Fig.~\ref{fig:power}, as well as of Fig.~\ref{fig:density}, where the relative magnetic flux averaged EP fluctuating distribution function, $\delta n/n_0$, is shown vs $(r/a)$, after integration in $\mu$ and over a narrow interval $\Delta v_\parallel$ near resonance. From these results, it is clear that the wave interaction with co-passing particles is lasting longer and covering a broader radial domain, of the order of the radial mode width.

\begin{figure}[t]
\includegraphics[width=75mm,height=35mm]{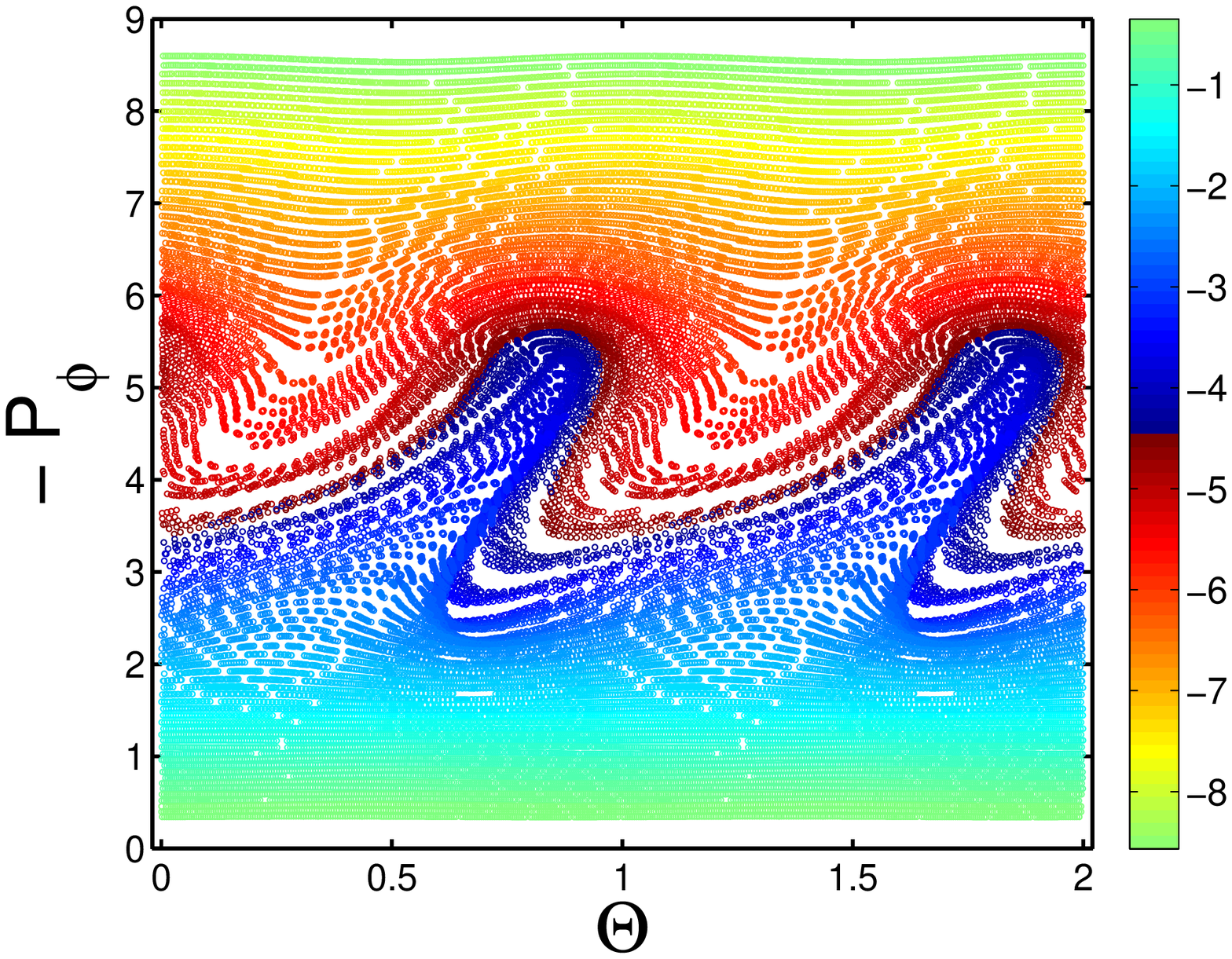}
\includegraphics[width=75mm,height=35mm]{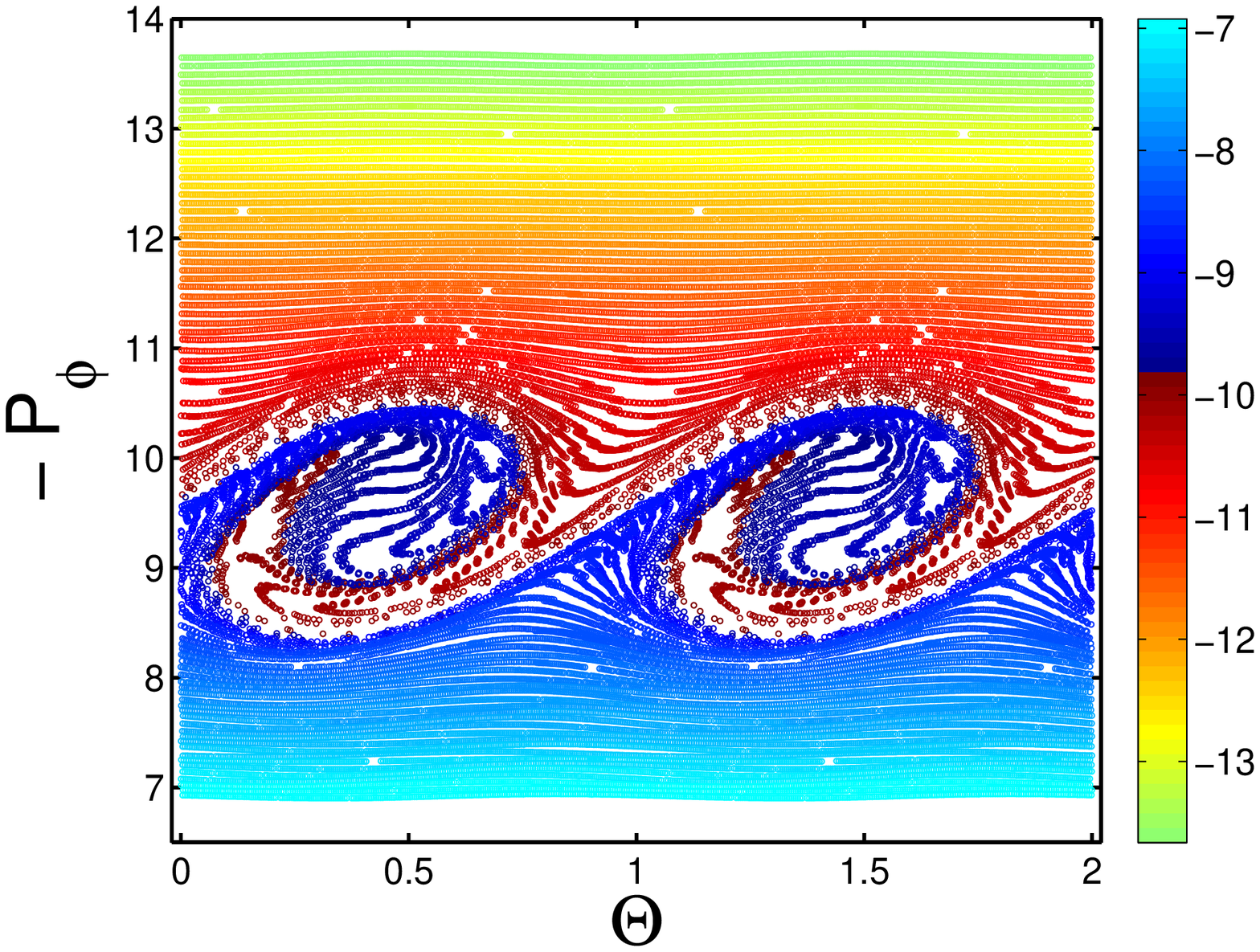}
\caption{\label{fig:phase} Phase space structures in $(\Theta, -P_\phi)$ ($\Theta$ and $P_\phi$ are normalized w.r.t. $2\pi$ and $a\sqrt{mE_0}$ with $a$ the minor radius) phase-space for co-passing particles (upper panel) and counter-passing particles (lower panel) at saturation, where $\Theta$ is sampled at $\theta=0$.}
\end{figure}

Equation~(\ref{eq:detuning}) also provides useful insights for the analysis of phase-space structures shown in Fig.~\ref{fig:phase}. There, test particle motion in the phase-space is illustrated for both co- and counter-passing particles. For $v_\parallel>0$, the wave-particle phase $\Theta$ is essentially constant, i.e. the $\dot{\Theta}\simeq0$ resonance condition is effectively maintained through saturation. This occurs because of Eq.~(\ref{eq:fldr}), showing that frequency chirps up due to the effect of EP radial transport. Frequency chirping, in turn, is connected with phase-locking $(\dot{\Theta}\simeq0)$, for this is the condition that allows particles to be most efficiently transported outward, while driving the mode. This condition is similar to that, which underlies EP transports in avalanches~\cite{zonca05}, due to EPM~\cite{chen94}. In the present case, however, the finite BAE radial structure determines the finite interaction length and mode saturation is reached when particles are pumped out of the effective mode radial domain. For counter-passing particles, meanwhile, Fig.~\ref{fig:phase} shows rapid wave-particle phase mixing, as expected from Eq.~(\ref{eq:detuning}), and the corresponding formation of phase-space island structures near resonance, which explain the reduced drive for $v_\parallel<0$ in Fig.~\ref{fig:power}~\cite{oneil65,mazitov65}. This result, which crucially depends on radial non-uniformity of the system and applies for $|\omega-\omega_0|\gtrsim\gamma_L$, is different from the well-known case of nonlinear saturation due to wave-particle trapping~\cite{onishchenko70,oneil71,shapiro71,berk92}. It is also different from the adiabatic frequency chirping of hole-clump structures in phase-space~\cite{berk97}, since, in the present case, $|\dot{\omega}|\gtrsim\gamma^2_L$ (c.f. Eq.~(\ref{eq:detuning})) and the frequency sweeping is non-adiabatic~\cite{chen07,zonca99}. This is a crucial element for maximizing EP transport, for a broader region of the phase-space can be affected by radial redistribution, as shown in Fig.~\ref{fig:density} and~\ref{fig:phase} by the $v_\parallel$ symmetry breaking. Meanwhile, considering Eq.~(\ref{eq:fldr}), this explains why frequency chirping is connected with the nonlinear evolution of the mode amplitude, as shown in Fig.~\ref{fig:omega} and often observed in experiments~\cite{podesta11}. 

\begin{figure}[t]
\includegraphics[width=80mm,height=49mm]{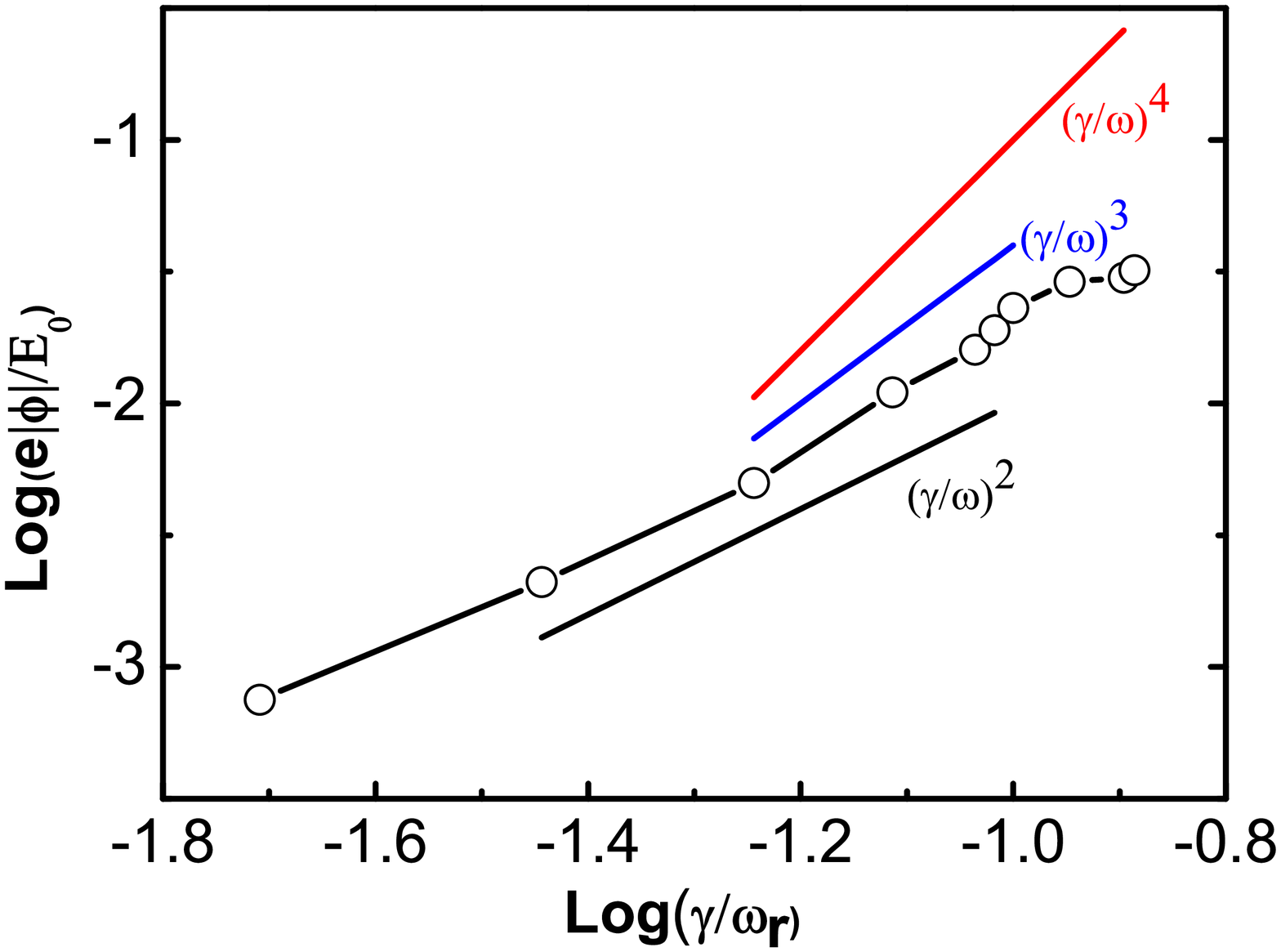}
\caption{\label{fig:amp} The saturation amplitude of the fluctuating scalar potential vs. the linear growth rate at different values of EP density. The "$\circ$" are simulation results. The black, blue and red lines are, respectively, denoting $\propto(\gamma/\omega)^2, (\gamma/\omega)^3, (\gamma/\omega)^4$ for references.  }
\end{figure}

This interpretation of simulation results is further supported by the scaling of the saturation amplitude with the growth rate, which is shown in Fig.~\ref{fig:amp} and can be understood theoretically by analyzing the nonlinear pendulum about the stable fixed point by radially localized perturbation. The corresponding wave-particle resonance is then described by
\begin{eqnarray}\label{eq:pendulum1}
dY/d\tau &=& X \nonumber \\
dX/d\tau &=& -A_0Y/(1+X^2) \ \ .
\end{eqnarray}
Here, $X=|k_r|x/(nq')$ and $Y=y|k_r|/(nq')$ are the rescaled variables for $x = nq'\Delta r$, $y = \Theta - \Theta_0$, $\Theta_0$ begin the fixed point and $\tau=\omega_t t$. We assume $A=A_0/(1+X^2)$ to investigate the effect of finite  characteristic radial mode width, i.e. $k^{-1}_r$, where $A_0$ is in general proportional to the mode saturation amplitude. Equation~(\ref{eq:pendulum1}) is separable and can be trivially integrated by quadratures. As a result, the period is given by ${\cal T} = 2\pi A_0^{-1/2}/\Gamma$, with
\begin{equation}\label{eq:int}
\sqrt{A_0} \Gamma = \left( \int^1_0\frac{(2/\pi)(1+z^2X_0^2)dz}{\sqrt{(1+X^2_0/2)-z^2(1+z^2X^2_0/2)}} \right)^{-1}  .
\end{equation}
The integral in Eq.~(\ref{eq:int}) is a function of $X_0$, the normalized maximum radial particle excursion, and is easily computed by asymptotic expansion, yielding $1$ for $X_0\rightarrow 0$ and $\propto X_0$ for $X_0\gg1$. The characteristic rate for a particle to undergo a radial excursion $X_0=|k_r|\Delta r_0$, is then given by $\gamma_{NL} =A_0^{1/2} \omega_t \Gamma(X_0)$.
If $X_0=|k_r|\Delta r_0\ll1$, as assumed in the nonlinear saturation of TAE modes~\cite{berk90}, the nonlinear dynamics is the same as in a uniform system and wave saturation can occur only when the particle distribution function in the resonance region is flattened, so that the nonlinear drive is significantly reduced and brought back to threshold. This occurs  for $\gamma_{NL}/\omega_t \sim \gamma_L/\omega_t\sim A_0^{1/2}$, yielding the well known estimate $A_0\propto(\gamma_L/\omega_t)^2$~\cite{onishchenko70,oneil71,shapiro71}.
When $X_0 = |k_r|\Delta r_0 \gtrsim 1$, saturation occurs because particle get off resonance by radial detuning, i.e., after a radial displacement of the order of the mode width, as noted above and observed in~\cite{briguglio98} for the first time. Similar to the uniform case, saturation is expected when $\gamma_{NL}/\omega_t   = A_0^{1/2} \Gamma(X_0) \propto A^{1/2}_0/X_0\approx\gamma_L/\omega_t$. Meanwhile, noting that $Y \sim |k_r|/(nq')$, Eqs.~(\ref{eq:pendulum1}) also give $X_0 \propto (A_0^{1/2}/X_0)|k_r|/(nq')$.  Thus, the saturation condition becomes $A_0\propto(\gamma_L/\omega)^4(k_r)^2/(nq')^2$.  By further increasing the drive, the radial scale $1/(nq')$ of the BAE mode will become increasingly more important, till the saturation amplitude is independent of $(\gamma_L/\omega)$.
For BAE modes, the transition between these two regimes
is expected to occur for $X_0\simeq 1$, i.e. for $A_0\approx(nq')^2/k^2_r$, where the radial wave number estimate for the BAE short scale radial structure is $|k_r|/(nq')\propto(\delta\omega/\omega)^{-1/2}$~\cite{zonca96a}, i.e. $|k_r|/(nq')\propto(\gamma_L/\omega)^{-1/2}$ for $\gamma_L\gtrsim|\mathbb{R}e\delta\omega|$. The expected scaling is then given by $A_0\propto(\gamma_L/\omega)^3$. Figure~\ref{fig:amp} illustrates the transition between these various regimes as $\gamma_L/\omega$ is increased, further demonstrating the important role of plasma non-uniformity and radial mode structures for mode saturation and EP transport processes.

In summary, we have found that nonlinear saturation of BAEs, driven by slowing down EPs via transit resonance, is characterized by upward frequency chirping and $v_\parallel$ symmetry breaking between co- and counter-passing particles, which can be understood as the the evidence of resonance-detuning. Upward frequency chirping is the preferred condition for maximizing EP radial transport while they drive the mode, and saturation eventually occurs when the particle radial displacement is of the order of the radial mode width. The scaling of the saturation amplitude with the growth rate further demonstrates the role of radial resonance detuning, due to the radial non-uniformity and mode structure. Both EP phase space behaviors as well as nonlinear dynamics and mode structures demonstrate the crucial roles of non-uniformities and geometry of the system. 

\begin{acknowledgments}
This work was supported by the National ITER Program of China and the Euratom Communities under the contract of Association between EURATOM/ENEA. One of the authors, Xin Wang, would like to thank R.B. White and H. Zhang for the valuable suggestions in the analysis of phase space structures; Liu Chen also acknowledges support of U.S. DoE Grant. Useful discussions with Z. Lin are also kindly acknowledged.  
\end{acknowledgments}

\end{document}